# A Gradual Takeover Strategy of the Active Safety System

Rui Liu[1], Xichan Zhu[2], Xuan Zhao[1*], Jian Ma[1]
1. School of Automobile, Chang'an University, Xi'an 710064, China
2. School of Automotive Studies, Tongji University, Shanghai 201804, China

**Abstract:** A gradual takeover strategy is proposed, in which the dynamic driving privilege assignment in real-time and the driving privilege gradual handover are realized. Firstly, the driving privilege assignment based on the risk level is achieved. The naturalistic driving data is applied to study the driver behavior during danger. TTC (time to collision) is defined as an obvious risk measure, whereas the time before the host vehicle has to brake assuming that the target vehicle is braking is defined as the potential risk measure, i.e. the time margin (TM). A risk assessment algorithm is proposed based on the obvious risk and potential risk. Secondly, the driving privilege gradual handover is realized. The non-cooperative MPC (model predictive control) is employed to resolve the conflicts between the driver and active safety system. The naturalistic driving data are applied to verify the effectiveness of the risk assessment algorithm, and the risk assessment algorithm performs better than TTC in the ROC (receiver operating characteristic). It is identified that the Nash equilibrium of the non-cooperative MPC can be achieved by using a non-iterative method. The driving privilege gradual handover is realized by using the confidence matrixes updating. The simulation verification shows that the gradual takeover strategy can achieve the driving privilege gradual handover between the driver and active safety system.

## 1. Introduction

Automotive intelligence has played an important role in reducing traffic accidents and reducing driver operating load[1]. The high-level automated driving system has made great progress in recent years. However, many accidents related to intelligent vehicle have shown that it is very important to keep the driver in the control loop before the driving automation system is fully mature[2, 3], i.e. driver assistance and cooperative driving.

The entire driving process can be divided into 4 states according to the risk level[4, 5], i.e. normal driving, risk, pre-crash, and crash. The boundary between the normal driving and the risk is the subjective safety domain limit, which is related to the driver's subjective perception of danger. The boundary between the risk and the pre-crash is the physical safety domain limit, which is the last moment at which collision avoidance can be completed by braking through maximum braking deceleration or emergency steering with maximum peak lateral acceleration[6].

In the traditional active safety system, the warning system starts to work when encountering the risk state, and the AEB (autonomous emergency braking) starts to work when entering the pre-crash state. However, the development of the danger is a gradual process. The suddenly intervention of the AEB will disturb the driver, and the intervention strategy of AEB is usually very conservative. Moreover, it will be difficult for the driver to take over when the driving automation system suddenly withdraws. If the active safety system can gradually intervene as the risk level increases, it will greatly increase the safety benefits of the active safety system.

When the driving privilege is delivered gradually, the vehicle is jointly controlled by the driver and active safety system, i.e. cooperative driving. The prerequisite for the realization of gradual takeover is risk assessment. The driving privilege is assigned between the driver and active safety system according to the risk level. Driver's subjective judgment of danger is particularly important for the definition of the risk level, i.e. the subjective safety

Corresponding Author: Prof. Xuan Zhao. Address: No. 33, Middle Chang'an Road, Xi'an 710064, China. E-mail: zhaoxuan@chd.edu.cn. Phone: (+86) 13572219879.

domain. In order to accurately define the boundary of the subjective safety domain, it is necessary to study the driving behavior characteristics of the driver in dangerous scenario. Hence, the Naturalistic Driving Data (NDD) is applied to achieve the driver behavior during danger.

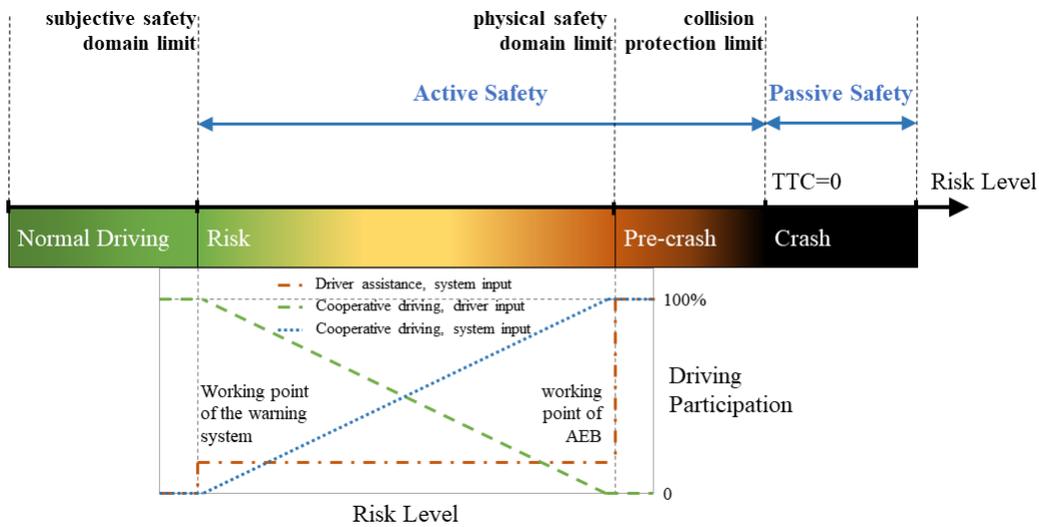

Fig 1. Gradual takeover strategy during danger

The common risk assessment method is to define a measure that characterizes the risk level in each scenario. TTC (time to collision) is a widely used risk measure[7-10], which is defined as the relative distance divided by the relative velocity. When the relative velocity is very small, TTC will be infinity. Hence, the reciprocal of TTC (1/TTC) is also used in risk assessment[11, 12]. However, TTC or 1/TTC cannot represent the risk level in all circumstances. Let's consider 2 car-following cases. In case 1, the velocity of the host vehicle and target vehicle are 31m/s and 30m/s, respectively. And the relative distance is 10m. In case 2, the velocity of the host vehicle and target vehicle are 6m/s and 5m/s, respectively. And the relative distance is 10m. The TTC in these two cases are all 10s, but the driver will feel more dangerous in case 1. If the relative distance is too small, the braking of the target vehicle will cause a rear-end collision risk, even if TTC is very large. Too small relative distance and insufficient braking force are the main factors leading to danger in the car-following scenario[13]. Therefore, TTC cannot be used to evaluate the risk level when the relative velocity is small. THW (time headway) is also a commonly used risk measure[14], which is defined as the relative distance divided by the velocity of the host vehicle. However, the velocity of the target vehicle is not considered in THW, which is easy to obtain and is very important in risk assessment. Furthermore, the driver's choice of THW is affected by many other factors including the target vehicle type and road condition[15]. Hence, THW also cannot describe the danger in car-following accurately. Some new measure are introduced to describe the risk level, e.g. the weighted sum of 1/TTC and 1/THW[16], $T_{lab}$[17].

The driving intention during driving privilege gradual handover come from the driver and active safety system. Therefore, the vehicle control according to the intentions of the two has become a problem. The non-cooperative dynamic game can be used to describe the problem that multiple decision makers act on the same dynamic system[18]. Hence, the non-cooperative dynamic game can be applied to deal with the conflict during driving privilege handover. The application of dynamic game in vehicle system is relatively limited. The dynamic game is employed to study the vehicle evaluation method under worst-case[19, 20]. The vehicle control strategies which take the driver behavior in to consideration are constructed based on the closed-loop game[21, 22]. The game theory is used to model the control behavior of the driver and the active front steering controller[23].

## 2. Naturalistic Driving Data

The natural driving data used in this paper come from 3 databases, i.e. the China-FOT (China field operational test), the dangerous scenario database, and the OEM FOT.

2.1 China-FOT

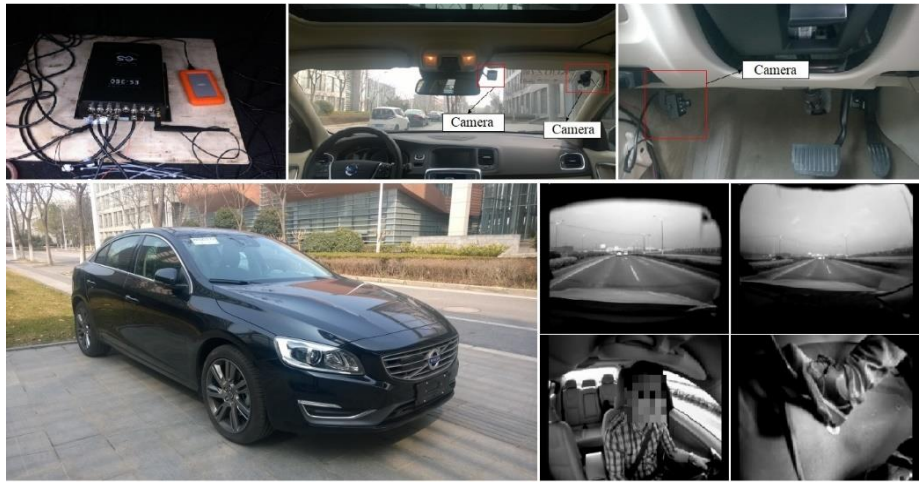

Fig 2. The test vehicle, data acquisition system and camera in the China-FOT

China-FOT is collected by using 8 test vehicles. The test vehicles are all Volvo S60L. The vehicle state information come from the CAN bus, and the surrounding traffic information are obtained by 4 cameras. Thirty-two drivers participate in the test, including 25 men and 7 women. The age of driver is between 28 and 39 (mean: 32.25; $SD$: 2.84). All drivers have their own vehicles before the test. The mileage ranges from 15,000km to 240,000km (mean: 108,375; $SD$: 63,598). Hence, all drivers in China-FOT are not newbie. Each driver uses the test vehicle for about 3 months. Drivers can drive the test car to any place at any time during the test. China-FOT have collected 7,402 trips. The travel distance is 129,935 km.

2.2 Dangerous Scenario Database

The dangerous scenario database is collected by using video drive recorder (VDR) installed on the vehicle. The VDR of Horiba with built-in velocity sensor and acceleration sensor is used. Brake deceleration equal to 0.4g is chosen as a trigger value, and the VDR only records the data within the period from 15s before to 5s after a trigger. About 4,000 trigger cases are collected during 4 years. The minimum TTC less than 2s is used to select the dangerous cases. And 1200 cases are obtained. The 1200 cases are manually screened, and the cases which may not be dangerous are eliminated. In the end, 500 dangerous cases with high risk level are obtained.

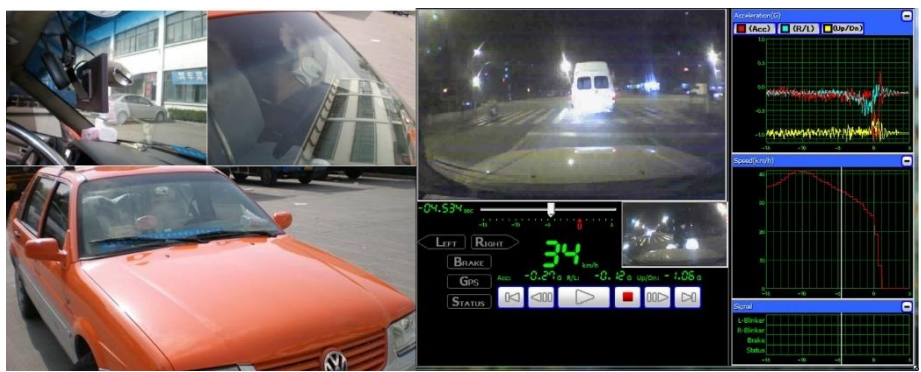

Fig 3. The test vehicle, VDR and data processing system in the dangerous scenario database

2.3 OEM FOT

The OEM FOT is derived from the field operational test of an anonymous OEM (original equipment manufacturer). All test vehicles are production passenger cars of the OEM. The vehicle state information come from the CAN bus, and the surrounding traffic information are obtained by using Mobieye EyeQ3. The Mobieye EyeQ3 provides information of 5 targets in front, including target type, target width, relative distance, relative speed, relative acceleration, etc. Only part of the data in the OEM FOT are available. The travel distance is 1,220 km.

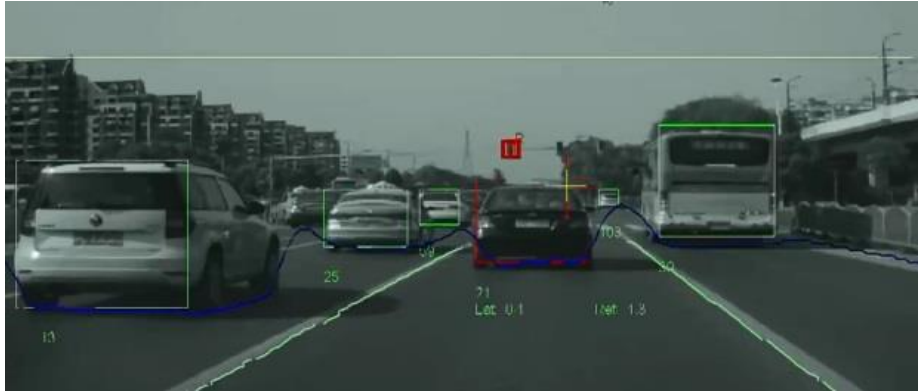

Fig 4. The Mobileye EyeQ3 video information in the OEM FOT

2.4 Scenario Extraction

Each of the 3 databases has its own advantages and disadvantages. The driving process of the driver is completely undisturbed in China-FOT. And China-FOT has long acquisition time, rich driving scenarios, and complete vehicle state information and video information. However, the surrounding traffic environment parameters in China-FOT are obtained through image identification. Hence, the accuracy is not high. The dangerous scenario database mainly collects dangerous cases. The risk level of the dangerous case is high. However, the dangerous scenario database only records data within 20s for each case. The OEM FOT has high quality surrounding traffic environment information, but the amount of data is limited. Therefore, different scenarios are extracted by using different databases according to the features of these 3 databases.

a) The cut-in cases are extracted by using China-FOT, and 326 cut-in cases are obtained. These cut-in cases are classified into normal cut-in and dangerous cut-in by using the automatic detection method introduced in [24]. In the last, 249 normal cut-in cases and 77 dangerous cut-in cases are obtained.

b) The dangerous car-following cases are extracted by using the dangerous scenario database. The 500 dangerous cases are classified, and 75 dangerous car-following cases with high risk level are obtained.

c) The normal car-following cases are extracted by using the OEM FOT. This FOT is a control test for certain ADAS functions. Hence, the ADAS is working at certain times during the test. Firstly, the manual driving data are picked out by using the variables in the CAN bus which indicate the on/off of the ADAS. Next, the car-following cases are extracted by using the information collected by the Mobieye EyeQ3, including the target type, lateral and longitudinal relative distance, longitudinal relative velocity. In the last, manual screening is applied to remove the car-following cases which may be dangerous. And 822 normal car-following cases are obtained.

**3. Driving Privilege Assignment based on Risk Level**

3.1 Obvious Risk and Potential Risk

As have mentioned before, TTC or 1/TTC cannot accurately represent the risk level when the relative velocity is small. It will be helpful to take the possible future braking operations of the target vehicle into account, e.g. the Mazda avoidance logic[25], the safety margin[26], and the responsibility sensitive safety (RSS)[27]. In this paper, TTC is defined as an obvious risk measure, whereas the time before the host vehicle has to brake assuming that the target vehicle is braking is defined as the potential risk measure, i.e. the time margin (TM).

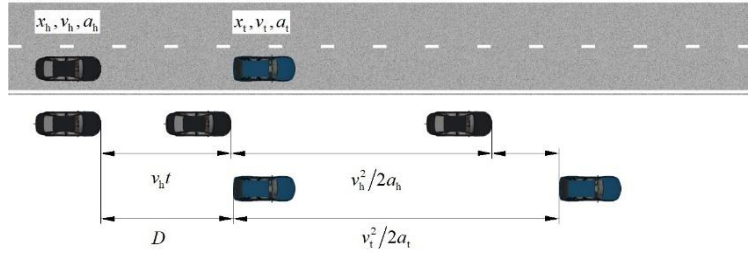

Fig 5. The collision avoidance process

The collision avoidance process when the target vehicle brakes to stop with a constant deceleration is shown in Fig 5. $x_h$, $v_h$, $a_h$ are the position, velocity, and brake deceleration of the host vehicle, respectively. $x_t$, $v_t$, and $a_t$ are the position, velocity, and brake deceleration of the target vehicle, respectively. $a_h$ and $a_t$ take the absolute value of the brake deceleration. $D$ is the relative distance. $t$ is the time from the current time until the host vehicle starts to decelerate. $t$ should contain 3 parts, i.e. the driver braking reaction time $\tau_1$, the braking system reaction time $\tau_2$, and the time $t_0$ which the driver can freely use. In order to avoid the collision when the target vehicle brakes, it should be

$$D + v_t^2/2a_t \geq v_h t + v_h^2/2a_h \qquad (1)$$

That is

$$t_0 + \tau_1 + \tau_2 \leq \frac{D + v_t^2/2a_t - v_h^2/2a_h}{v_h} \qquad (2)$$

When the target vehicle brakes with a constant deceleration, the maximum value of the sum of $\tau_1$, $\tau_2$ and $t_0$ is defined as the time margin (TM), i.e.

$$\text{TM} = \frac{D + v_t^2/2a_t - v_h^2/2a_h}{v_h} \qquad (3)$$

The brake deceleration of the host vehicle and the target vehicle is selected according to the friction limit of the vehicle, i.e. $a_h = a_t = 7 m/s^2$. TM indicates the reaction time left to the driver of the host vehicle if the target vehicle starts to brake. TTC indicates the risk level in the current state, whereas TM indicates the risk level if the target vehicle suddenly brakes. Therefore, TTC is defined as a obvious risk measure, and TM is defined as a potential risk measure. TM is mainly used to characterize the risk level when the relative velocity is small.

3.2 Risk Assessment in Car-following

A risk assessment algorithm is proposed by using the 75 dangerous car-following cases. In the dangerous car-following cases, the moment when the vehicle starts to decelerate is the time when the driver feels the danger and reacts to the it. And this moment is defined as the break starting time. Therefore, the dangerous threshold is determined by using the TTC and TM at the braking starting time.

Since TTC may become very large when the risk level is low, 1/TTC is applied to define the obvious risk level. The 1/TTC at the braking start time is used to determine the obvious risk threshold. It is found that the 1/TTC at last-second braking onset is related to the velocity of the host vehicle[28]. Hence, it is necessary to discuss whether the velocity of the host vehicle has a significant influence on the 1/TTC at the braking start time. The relationship between the 1/TTC and velocity is shown in Fig 6. The regression coefficient test is used to verify whether the 1/TTC has a significant regression relationship with the velocity. The results are shown in Table 1. The Durbin-Watson test indicates that the residual has no significant autocorrelation. And the data is suitable for regression analysis. The regression coefficient test shows that 1/TTC and velocity have a significant regression relationship. Therefore, the impact of the velocity should be considered when 1/TTC is used to classify the risk level. The empirical regression coefficient between 1/TTC and velocity is -0.0717.

Table 1 The results of regression coefficient test

|       | Durbin-Watson test | | Regression coefficient test | |
|---|---|---|---|---|
|       | $D.W.$ value | $p$ value | $t$ value | $p$ value |
| 1/TTC | 2.022 | $p<0.01$ | $t(73)=-3.850$ | $p<0.001$ |
| TM    | 1.867 | $p<0.01$ | $t(73)=-1.193$ | $p=0.237$ |

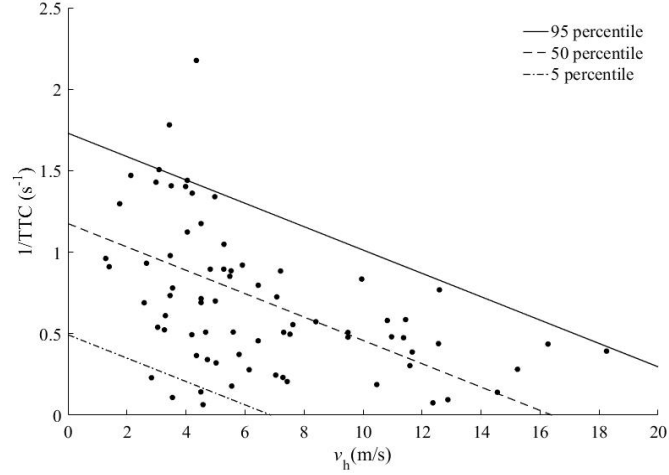

Fig 6. The 1/TTC at the brake starting time and the velocity of the host vehicle

The 5th, 50th, and 95th percentiles of the 1/TTC at the braking start time in 75 dangerous car-following cases are

$$\begin{cases} ittc_{95} = -0.0717v_x + 1.73 \\ ittc_{50} = -0.0717v_x + 1.18 \\ ittc_5 = -0.0717v_x + 0.49 \end{cases} \quad (4)$$

These 3 percentiles indicate that 5%, 50%, and 95% of the drivers brake when the 1/TTC reaches the corresponding threshold. When the obvious risk level is divided by using the threshold associated with the velocity, the thresholds will reach 0 as the velocity increases. A minimum value of the 1/TTC is set for each obvious risk level. Consequently, the obvious risk level is

$$\begin{cases} OR0: 1/TTC < \max(ittc_5, thr_1); \\ OR1: \max(ittc_5, thr_1) \leq 1/TTC < \max(ittc_{50}, thr_2); \\ OR2: \max(ittc_{50}, thr_2) \leq 1/TTC < \max(ittc_{95}, thr_3); \\ OR3: 1/TTC \geq \max(ittc_{95}, thr_3). \end{cases} \quad (5)$$

Where OR0 means no obvious risk. OR1, OR2 and OR3 indicate the level 1, level 2 and level 3 obvious risk level. $thr_1$, $thr_2$, and $thr_3$ are the 1/TTC minimum values in the 3 risk levels, respectively. Referring to [29], the 1/TTC minimum values are set as $thr_1=0.33s^{-1}$, $thr_2=0.66s^{-1}$, and $thr_3=1s^{-1}$.

The TM at the brake starting time is used to determine the potential risk threshold. Similarly, the regression coefficient hypothesis test is used to discuss whether TM has a significant regression relationship with the velocity. The relationship between the TM and velocity is shown in Fig 7. The regression coefficient test is used to verify whether TM has a significant regression relationship with the velocity. The results are shown in Table 1. The Durbin-Watson test indicates that the residual has no significant autocorrelation. And the data is suitable for regression analysis. The regression coefficient test shows that TM and velocity have no significant regression relationship. Therefore, the influence of the velocity is not considered in the potential risk level. And the horizontal lines are employed to divide the TM thresholds.

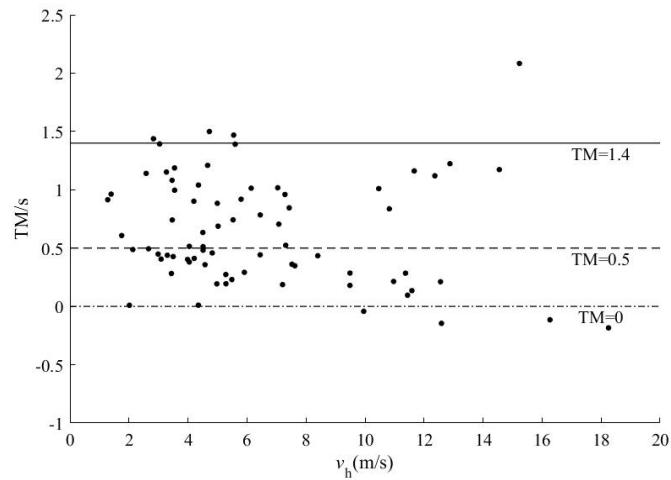

Fig 7. The TM at the brake starting time and the velocity of the host vehicle

The 5th, 50th, and 95th percentiles of TM at the brake starting time in 75 dangerous car-following cases are 1.4s, 0.5s, and 0, respectively. These 3 percentiles indicate that 5%, 50%, and 95% of the drivers brake when TM reaches the corresponding threshold. These 3 percentiles are applied to be the thresholds for the potential risk level, i.e.

$$\begin{cases} PR0: TM > 1.4s; \\ PR1: 0.5s < TM \leq 1.4s; \\ PR2: 0 < TM \leq 0.5s; \\ PR3: TM \leq 0. \end{cases} \quad (6)$$

Where PR0 means no potential risk. PR1, PR2 and PR3 represent the level 1, level 2 and level 3 potential risk level, respectively. Note that when TM<0, it means that the host vehicle cannot avoid a collision if the target vehicle suddenly brakes at the maximum deceleration. Hence, TM<0 is reasonable in some cases.

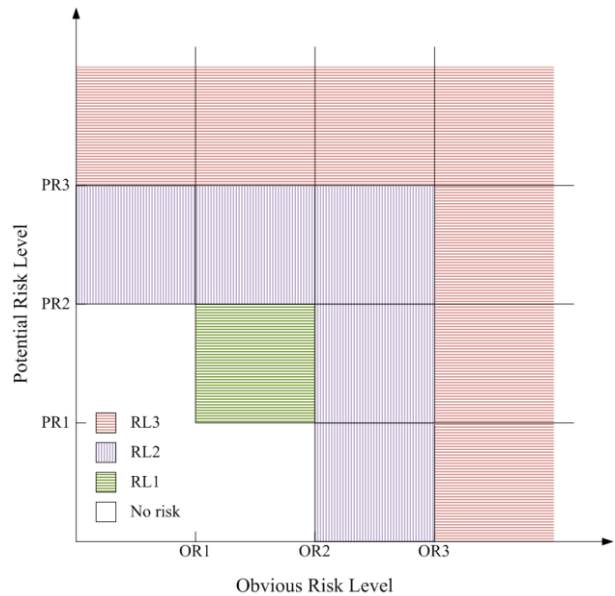

Fig 8. The risk assessment algorithm

The risk assessment algorithm which considers the obvious risk and potential risk is

$$\begin{cases} \text{RL0: (PR0\&\&OR0)||(PR1\&\&OR0)||(PR0\&\&OR1)} \\ \text{RL1: PR1\&\&OR1} \\ \text{RL2: (PR2||OR2)\&\&(!PR3)\&\&(!OR3)} \\ \text{RL3: PR3||OR3} \end{cases} \quad (7)$$

Where RL0 means no risk. RL1, RL2 and RL3 are the risk level 1, risk level 2 and risk level 3, respectively. && indicates logical and; || indicates logical or; ! indicates logical not.

PR1 and OR1 are the thresholds after removing a small number of abnormal cases. If the driver is warned at PR1 or OR1, many false alarms will emerge. When PR1 and OR1 are simultaneously achieved, excessive false alarm can be avoided. Therefore, the RL1 is achieved when PR1 and OR1 are reached at the same time. PR3 and OR3 use the 95th percentile parameter, which is very urgent danger. Hence, RL3 is achieved when one of PR3 or OR3 is reached. The other situation between RL1 and RL3 is set as RL2.

Four examples in the 75 dangerous car-following cases are picked out to demonstrate the evolution of obvious risk and potential risk in the car-following cases. The velocity of the host vehicle ($v_h$), the relative velocity ($v_r$), TM and 1/TTC of the 4 cases are shown in Fig 9 to Fig 12. Fig 9 is a danger that occurs shortly after the host vehicle starts. Fig 10 is a danger when the host vehicle approaches a slowly moving target vehicle. The relative velocity is large in these 2 cases at the beginning. 1/TTC can detect the danger in these cases which have high relative velocity. However, the risk assessment algorithm still can detect the danger earlier when the relative velocity is high. When a potential risk emerges, the risk assessment algorithm can determine the danger when the TTC is about 3s. If the potential risk is not considered, it will cause a lot of false alarms if the moment when TTC equals to 3s is judged as a danger. TM can assist 1/TTC in defining the warning or intervention moment more precisely. Fig 11 is a stable car-following case in the urban elevated road. Fig 12 is a stable car-following case in the city road. The relative speed is very small in these 2 cases at the beginning, and the 1/TTC is also very small when the relative velocity is low. The danger is caused by the sudden braking of the target vehicle. However, TM has reached PR1 more than 10s before the driver of the host vehicle brakes. This indicates that the relative distance is too small, though the 1/TTC is very small. In the cases whose relative distance is too small, the obvious risk level will increase rapidly if the target vehicle suddenly brakes. In the stable car-following case, the potential risk can help to detect the danger much earlier. TM makes up for the shortcomings of TTC that cannot describe the danger accurately in the cases with small relative velocity. Therefore, the risk estimation algorithm that considers both obvious risk and potential risk is better than TTC.

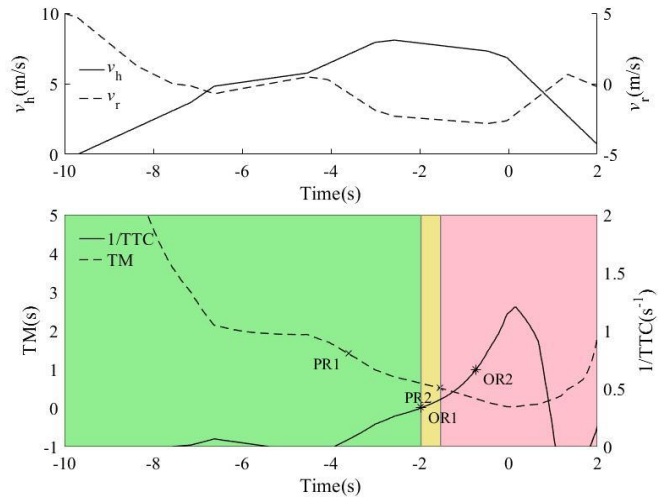

Fig 9. The velocity, TM and 1/TTC in the first car-following case

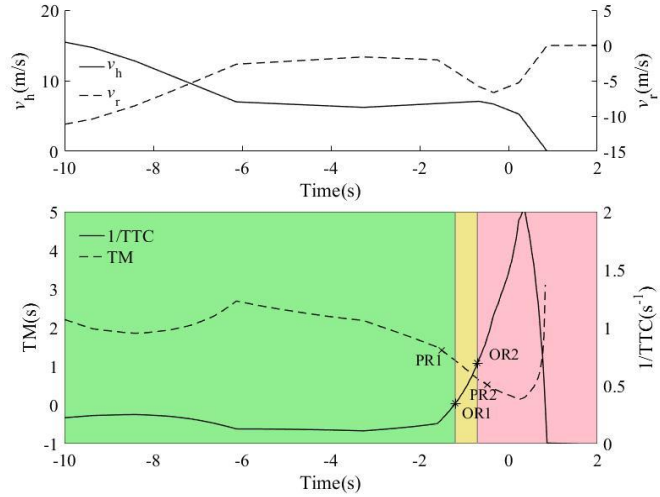

Fig 10. The velocity, TM and 1/TTC in the second car-following case

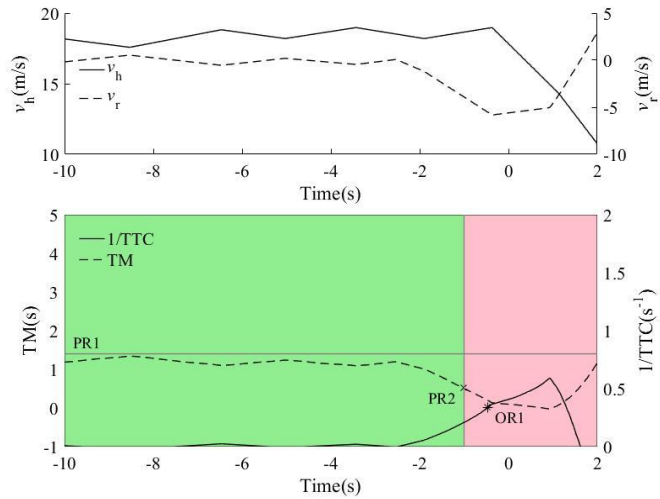

Fig 11. The velocity, TM and 1/TTC in the third car-following case

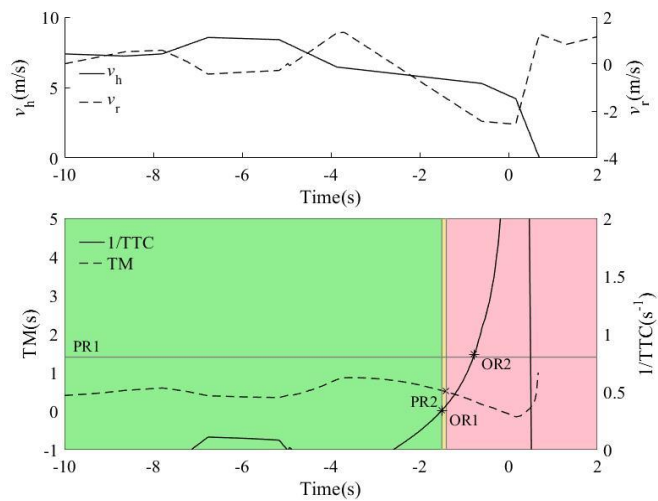

Fig 12. The velocity, TM and 1/TTC in the fourth car-following case

The confusion matrix[30] are used to evaluate the effectiveness of the risk assessment algorithm. Many evaluation indicators can be obtained based on the confusion matrix, as shown in Table 3. The *accuracy* is a good evaluation indicator when the number of "positive" and "negative" are similar. When one type of data accounts for the majority, the *accuracy* will mainly consider the classification accuracy of the majority, and the classification accuracy of the minority will not have a significant impact on the result. The receiver operating characteristic (ROC) is not sensitive to data proportion[31]. Hence, the ROC is applied to compare the effectiveness of the TTC, TM, and risk assessment algorithms.

Table 2. Confusion matrix

|  |  | Actual State | |
|---|---|---|---|
|  |  | Positive | Negative |
| Detection | Positive | True Positive (*TP*) | False Positive (*FP*) |
| State | Negative | False Negative (*FN*) | True Negative (*TN*) |

Table 3. Evaluation index based on confusion matrix

| Index | Definition | Index | Definition |
|---|---|---|---|
| *TP* rate/*sensitivity* | $\dfrac{TP}{TP+FN}$ | *FP* rate | $\dfrac{FP}{FP+TN}$ |
| *FN* rate | $\dfrac{FN}{TP+FN}$ | *Accuracy* | $\dfrac{TP+TN}{TP+FN+FP+TN}$ |
| *TN* rate | $\dfrac{TN}{FP+TN}$ | *Precision* | $\dfrac{TP}{TP+FP}$ |

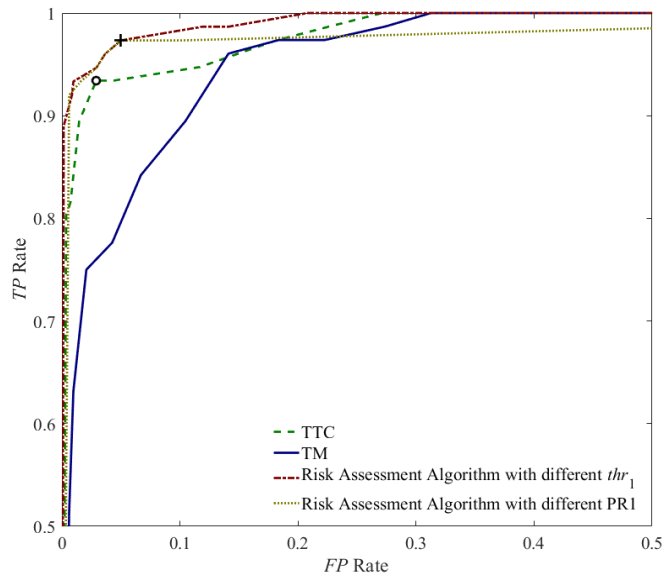

Fig 13. The ROC of the TTC, TM and risk assessment algorithm

The *TP* rate of TTC and TM are verified by using the 75 dangerous car-following cases, and the *FP* rate of TTC and TM are verified by using the 822 normal car-following cases. The ROC curves of TTC and TM are shown in Fig 13. The *TP* rate and *FP* rate of the risk assessment algorithm are marked in Fig 13 with a symbol '+'. The value of $thr_1$ and the thresholds of PR1 have enormous effect on the *TP* rate and *FP* rate of the risk assessment algorithm. Hence, Fig 13 demonstrates the ROC curve of the risk assessment algorithm with different $thr_1$ when PR1 threshold is TM=1.4s and the ROC curve of the risk assessment algorithm with different PR1 thresholds when $thr_1=0.33s^{-1}$. The ROC curve of TTC is completely contained inside the ROC curve of the risk assessment algorithm when $thr_1$ is different, which indicates that the risk assessment algorithm is always better than TTC.

The ROC curve of TTC is very close to that of the risk estimation algorithm at one point with a symbol 'o'. But the TTC threshold for danger detection is 1.9s at this point. The danger is too late to be detected. The danger detection is much easier and the significance of the danger detection has been greatly reduced when the risk level is high. In addition to the better ROC performance, the 4 examples in Fig 9 to Fig 12 show that the risk estimation algorithm has better risk judgment timing. The *TP* rate of the risk assessment algorithm cannot reach 1 when PR1 is different. This indicates that TTC have more influence on *TP* rate rather than TM. And TM mainly help with reducing the *FP* rate.

3.3 Risk Assessment in Cut-in

The driving behavior of the driver in cut-in scenario is studied by using the 249 normal cut-in cases and 77 dangerous cut-in cases. Similarly, the time when the driver starts braking in the cut-in is the moment when the driver feels the danger and response to the danger. Therefore, the driver behavior at the brake starting time in the cut-in scenario is presented. The brake starting time is defined by the moment when the driver of the host vehicle steps the brake pedal, which can be distinguished by the video and the brake pressure in the CAN bus.

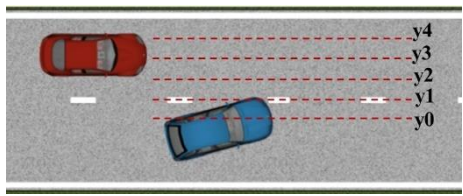

Fig 14. The lateral division of the lane

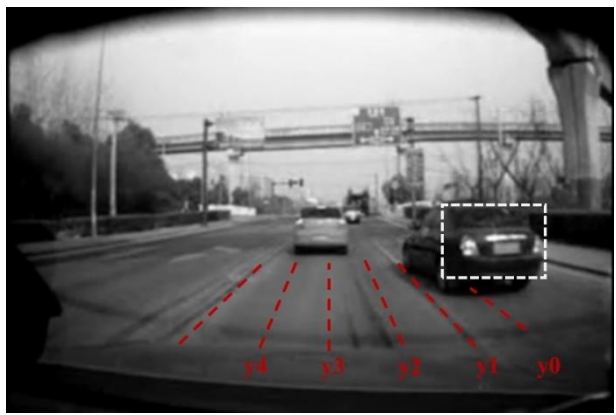

Fig 15. The lateral position of the target vehicle at brake starting time

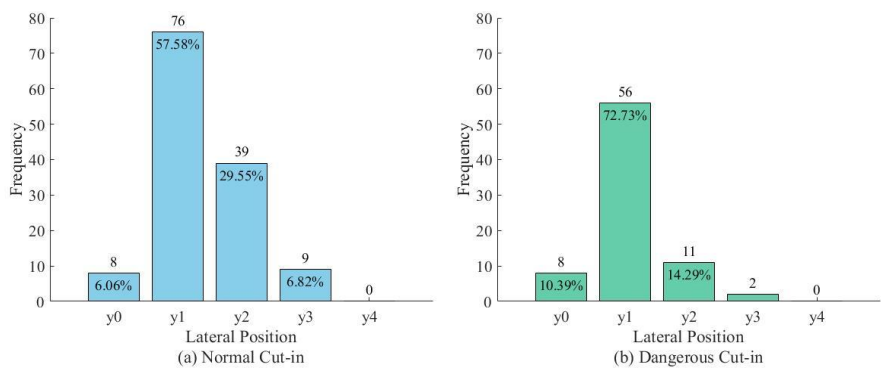

Fig 16. The lateral position in the cut-in cases

The lane is divided into five sections laterally, and the dividing lines are represented by y0 to y4, respectively. The distance between each dividing line is 1/4 lane width. The lateral position of the target vehicle at the brake starting time is analyzed. An example of the cut-in case at brake starting time is given in Fig 15, in which the target vehicle is at the position y1. The lateral position of the target vehicle in the 249 normal cut-in cases and 77 dangerous cut-in cases are shown in Fig 16. In dangerous cut-in cases, the brake time of the host vehicle is earlier than that in normal cut-in cases. In both normal cut-in cases and dangerous cut-in cases, most drivers will start braking when the target car reaches the lane line. Very few drivers start braking before the target car reaches the lane line. Therefore, the position that the target vehicle arrives at the lane line is chosen as the moment to start the risk assessment in the cut-in scenario. After the target vehicle reaches the lane line, the risk level is estimated by using the risk assessment algorithm in the car-following scenario.

3.4 Driving Privilege Assignment

The credibility of the driver's operation is related to the risk level. When the vehicle is in the normal driving state, it indicates that the driver can make a correct judgment on the traffic environment and maintain good control of the vehicle. The driver's operation is highly reliable at this time. When the vehicle enters a dangerous state, it indicates that the driver already has a misjudgment of the traffic environment or has incorrect operation. The credibility of the driver's operation is low at this time. When the vehicle enters a dangerous state from normal driving state, the driving privilege is gradually transferred from the driver to the active safety system. When the vehicle returns from a dangerous state to the normal driving state, the driving privilege is gradually returned from the active safety system to the driver. The weight factor of the driver is denoted as $\kappa_1$, and the weight factor of the active safety system is denoted as $\kappa_2$. The weight factors vary according to a linear law in the transfer of the driving privilege. The weight factors of the driver and the active safety system during the driving privilege handover can be expressed as

$$\kappa_1(k) = \begin{cases} a, k < k_0 \\ \dfrac{\beta - a}{K}(k - k_0) + a, k_0 \leq k \leq k_T \\ \beta, k > k_T \end{cases} \quad (8)$$

$$\kappa_2(k) = \Lambda - \kappa_1(k)$$

Where $\Lambda$ is the total assignable driving privilege. $\alpha$ is the driving weight of the driver in the current state, and $\beta$ is the driving weight assigned to the driver after the handover. $\alpha, \beta \in [0, \Lambda]$. $k_0$ is the step that starts the driving privilege handover, and $k_T$ is the step at the end of the handover. $K$ is the total number of steps in the entire driving privilege handover, $K = k_T - k_0$. If the driving right handover duration is $t$ and the time of each step is $T$, then $K = t/T$.

Three handover strategies are set according to the risk. When the vehicle is in normal driving state, the driving right is completely allocated to the driver. When entering the risk level 1, the driving privilege is slowly transferred from the driver to the active safety system within 3s. When entering the risk level 2 The driving privilege is handed over from the driver to the active safety system within 1s. When entering the risk level 3, the driving privilege is handed over from the driver to the active safety system within 0.5s. That is

$$\begin{cases} RL1: \kappa_1(k_0) = \Lambda, \kappa_2(k_0) = 0, \kappa_1(k_T) = 0, \kappa_2(k_T) = \Lambda, T = 3\,\text{s}; \\ RL2: \kappa_1(k_0) = \Lambda, \kappa_2(k_0) = 0, \kappa_1(k_T) = 0, \kappa_2(k_T) = \Lambda, T = 1\,\text{s}; \\ RL3: \kappa_1(k_0) = \Lambda, \kappa_2(k_0) = 0, \kappa_1(k_T) = 0, \kappa_2(k_T) = \Lambda, T = 0.5\,\text{s}; \end{cases} \quad (9)$$

When the vehicle completes the collision avoidance, the driver needs to be reminded and the driving privilege is returned to the driver from the active safety system. Two handover modes are set according to the state of the driver. If the driver has the intention to take over, the driving privilege is returned to the driver within 2s. If the driver does not have the intention to take over, the driving privilege is returned to the driver within 6s. That is

$$\begin{cases} \text{Mode 1}: \kappa_1(k_0) = 0, \kappa_2(k_0) = \Lambda, \kappa_1(k_T) = \Lambda, \kappa_2(k_T) = 0, T = 2\,\text{s}; \\ \text{Mode 2}: \kappa_1(k_0) = 0, \kappa_2(k_0) = \Lambda, \kappa_1(k_T) = \Lambda, \kappa_2(k_T) = 0, T = 6\,\text{s}; \end{cases} \quad (10)$$

The takeover intention and takeover state of the driver can be obtained through DMS (driver monitoring system).

# 4. Driving Privilege Gradual Handover based on Non-cooperative MPC

## 4.1 System Model

The joint control system model in the gradual takeover can be expressed as

$$Z_f(k) = \Psi_f x_f(k) + \Theta_{f,1} U_{f,1}(k) + \Theta_{f,2} U_{f,2}(k) + \Xi_f D_f(k) \tag{11}$$
$$f \in \{x, y\}$$

With

$$Z_f(k) = \begin{bmatrix} z_f(k+1) \\ z_f(k+2) \\ \vdots \\ z_f(k+N_p-1) \\ z_f(k+N_p) \end{bmatrix}, \quad U_{f,1} = \begin{bmatrix} u_{f,D}(k) \\ u_{f,D}(k+1) \\ \vdots \\ u_{f,D}(k+N_u-2) \\ u_{f,D}(k+N_u-1) \end{bmatrix}, \quad U_{f,2} = \begin{bmatrix} u_{f,A}(k) \\ u_{f,A}(k+1) \\ \vdots \\ u_{f,A}(k+N_u-2) \\ u_{f,A}(k+N_u-1) \end{bmatrix}, \quad \Psi_f = \begin{bmatrix} C_f A_f \\ C_f A_f^2 \\ \vdots \\ C_f A_f^{N_p-1} \\ C_f A_f^{N_p} \end{bmatrix},$$

$$D_f(k) = \begin{bmatrix} w_f(k) \\ w_f(k+1) \\ \vdots \\ w_f(k+N_u-2) \\ w_f(k+N_u-1) \end{bmatrix}, \quad \Xi_f = \begin{bmatrix} C_f B_{f,w} & 0 & \cdots & 0 \\ C_f A_f B_{f,w} & C_f B_{f,w} & \cdots & 0 \\ \vdots & \vdots & \ddots & \vdots \\ C_f A_f^{N_u-1} B_{f,w} & C_f A_f^{N_u-2} B_{f,w} & \cdots & C_f B_{f,w} \\ \vdots & \vdots & \ddots & \vdots \\ C_f A_f^{N_p-1} B_{f,w} & C_f A_f^{N_p-2} B_{f,w} & \cdots & C_f A_f^{N_p-N_u} B_{f,w} \end{bmatrix},$$

$$\Theta_{f,1} = \begin{bmatrix} C_f B_{f,1} & 0 & \cdots & 0 \\ C_f A_f B_{f,1} & C_f B_{f,1} & \cdots & 0 \\ \vdots & \vdots & \ddots & \vdots \\ C_f A_f^{N_u-1} B_{f,1} & C_f A_f^{N_u-2} B_{f,1} & \cdots & C_f B_{f,1} \\ \vdots & \vdots & \ddots & \vdots \\ C_f A_f^{N_p-1} B_{f,1} & C_f A_f^{N_p-2} B_{f,1} & \cdots & C_f A_f^{N_p-N_u} B_{f,1} \end{bmatrix},$$

$$\Theta_{f,2} = \begin{bmatrix} C_f B_{f,2} & 0 & \cdots & 0 \\ C_f A_f B_{f,2} & C_f B_{f,2} & \cdots & 0 \\ \vdots & \vdots & \ddots & \vdots \\ C_f A_f^{N_u-1} B_{f,2} & C_f A_f^{N_u-2} B_{f,2} & \cdots & C_f B_{f,2} \\ \vdots & \vdots & \ddots & \vdots \\ C_f A_f^{N_p-1} B_{f,2} & C_f A_f^{N_p-2} B_{f,2} & \cdots & C_f A_f^{N_p-N_u} B_{f,2} \end{bmatrix}.$$

Where $f=x$ denotes the longitudinal control model, and $f=y$ denotes the lateral control model. $x_f$ is the state variable, and $z_f$ is the observable state variable. $u_{f,D}$ is the input of the driver, and $u_{f,A}$ is the input of the active safety system. $A_f$, $B_{f,1}$, $B_{f,2}$, $C_f$ and $B_{f,w}$ are the constant matrixes. $w_f$ is the measurable disturbance. $N_p$ is the preview horizon. $N_u$ is the control horizon.

In the joint lateral control model of the vehicle steering, the lateral control state variable is $x_y=[d_y, v_y, \psi, \omega]^T$. Where $d_y$ is the lateral displacement, $v_y$ is the lateral velocity, $\psi$ is the yaw angle, and $\omega$ is the yaw rate. The state equation of lateral control can be expressed as

$$\dot{x}_y = A_{y,c} x_y + B_{y,1,c} u_{y,D} + B_{y,2,c} u_{y,A} \tag{12}$$
$$z_y = C_y x_y$$

With

$$A_{y,c} = \begin{bmatrix} 0 & 1 & u & 0 \\ 0 & a_{11} & 0 & a_{12} \\ 0 & 0 & 0 & 1 \\ 0 & a_{21} & 0 & a_{22} \end{bmatrix}, \quad B_{y,1,c} = B_{y,2,c} = \begin{bmatrix} 0 \\ b_1 \\ 0 \\ b_2 \end{bmatrix}, \quad C_y = \begin{bmatrix} 1 & 0 & 0 & 0 \\ 0 & 0 & 1 & 0 \end{bmatrix}, \quad a_{11} = -\frac{2C_f + 2C_r}{mv_x},$$

$$a_{12} = -v_x - \frac{2aC_f - 2bC_r}{mv_x}, \quad a_{21} = -\frac{2aC_f - 2bC_r}{I_z v_x}, \quad a_{22} = -\frac{2a^2 C_f + 2b^2 C_r}{I_z v_x}, \quad b_1 = \frac{2C_f}{m}, \quad b_2 = \frac{2aC_f}{I_z}.$$

Where $v_x$ is the longitudinal velocity of the vehicle. $C_f$ and $C_r$ are the cornering stiffness of each of the front and rear tires. $a$ and $b$ are the distances of the front and rear axles from the center of gravity of the vehicle. $I_z$ is yaw inertia of the vehicle. $m$ is the vehicle mass. $u_{y,D}$ is the steering input of the driver, and $u_{y,A}$ is the steering input of the active safety system. $z_y$ is the observable state variable.

The continuous state equation is discretized. The discrete state equation is

$$\begin{aligned} x_y(k+1) &= A_y x_y(k) + B_{y,1} u_{y,D}(k) + B_{y,2} u_{y,A}(k) \\ z_y(k) &= C_y x_y(k) \end{aligned} \quad (13)$$

Where $A_y$ is the matrix corresponding to $A_{y,c}$ after discretization. $B_{y,1}$ and $B_{y,2}$ are the matrixes corresponding to $B_{y,1,c}$ and $B_{y,2,c}$ after discretization. The discrete state space equation is iterated and (11) can be achieved. The matrix $\Xi_y$ and variable $D_y(k)$ are all 0 in the joint control system model of vehicle lateral control.

In the joint longitudinal control model of the vehicle accelerating and braking, the longitudinal control state variable is $x_x = [d_x, v_x]^T$. Where $d_x$ is the relative distance between the host vehicle and the target vehicle. $v_x$ is the longitudinal velocity. The state equation of longitudinal control can be expressed as

$$\begin{aligned} \dot{x}_x &= A_{x,c} x_x + B_{x,1,c} u_{x,D} + B_{x,2,c} u_{x,A} + B_{x,w,c} w_x \\ z_x &= C_x x_x \end{aligned} \quad (14)$$

With

$$A_{x,c} = \begin{bmatrix} 0 & -1 \\ 0 & 0 \end{bmatrix}, \quad B_{x,1,c} = B_{x,2,c} = \begin{bmatrix} 0 \\ 1 \end{bmatrix}, \quad B_{x,w,c} = \begin{bmatrix} 1 \\ 0 \end{bmatrix}, \quad C_x = \begin{bmatrix} 1 & 0 \\ 0 & 1 \end{bmatrix}.$$

In the longitudinal control model, $w_x$ is the target vehicle velocity, i.e. $w_x = v_t$. $u_{x,D}$ is the longitudinal control input of the driver, and $u_{x,A}$ is the longitudinal control input of the active safety system.

The continuous state equation is discretized. The discrete state equation is

$$\begin{aligned} x_x(k+1) &= A_x x_x(k) + B_{x,1} u_{x,D}(k) + B_{x,2} u_{x,A}(k) + B_{x,w} w_x(k) \\ z_x(k) &= C_x x_x(k) \end{aligned} \quad (15)$$

Where $A_x$ is the matrix corresponding to $A_{x,c}$ after discretization. $B_{x,1}$ and $B_{x,2}$ are the matrixes corresponding to $B_{x,1,c}$ and $B_{x,2,c}$ after discretization. The discrete state space equation is iterated and (11) can be achieved.

4.2 Cost Function

The 2 players in the non-cooperative MPC (driver and active safety system) expect to minimize the cost functions for their own goals

$$\begin{aligned} &\min_{U_1} V_1(k), \quad \min_{U_2} V_2(k) \\ &\text{s.t.} \quad Z(k) = \Psi x(k) + \Theta_1 U_1(k) + \Theta_2 U_2(k) \end{aligned} \quad (16)$$

Where $V_1(k)$ is the cost function of the driver and $V_2(k)$ is the cost function of the active safety system.

The cost functions of the 2 players are defined as

$$\begin{aligned} V_1(k) &= \|Z(k) - T_1(k)\|^2_{Q_1(k)} + \|U_1(k)\|^2_{R_1} \\ V_2(k) &= \|Z(k) - T_2(k)\|^2_{Q_2(k)} + \|U_2(k)\|^2_{R_2} \end{aligned} \quad (17)$$

With

$$T_1(k) = \begin{bmatrix} t_1(k-N_p+1) \\ t_1(k-N_p+2) \\ \vdots \\ t_1(k-1) \\ t_1(k) \end{bmatrix}, \quad T_2(k) = \begin{bmatrix} t_2(k-N_p+1) \\ t_2(k-N_p+2) \\ \vdots \\ t_2(k-1) \\ t_2(k) \end{bmatrix}, \quad R_1 = \begin{bmatrix} r_1 & 0 & \cdots & 0 \\ 0 & r_1 & \cdots & 0 \\ \vdots & \vdots & \ddots & \vdots \\ 0 & 0 & \cdots & r_1 \end{bmatrix}, \quad R_2 = \begin{bmatrix} r_2 & 0 & \cdots & 0 \\ 0 & r_2 & \cdots & 0 \\ \vdots & \vdots & \ddots & \vdots \\ 0 & 0 & \cdots & r_2 \end{bmatrix},$$

$$Q_1(k) = \begin{bmatrix} q_1(k+1) & 0 & \cdots & 0 \\ 0 & q_1(k+2) & \cdots & 0 \\ \vdots & \vdots & \ddots & \vdots \\ 0 & 0 & \cdots & q_1(k+N_p) \end{bmatrix}, \quad Q_2(k) = \begin{bmatrix} q_2(k+1) & 0 & \cdots & 0 \\ 0 & q_2(k+2) & \cdots & 0 \\ \vdots & \vdots & \ddots & \vdots \\ 0 & 0 & \cdots & q_2(k+N_p) \end{bmatrix},$$

$$q_1(k) = \begin{bmatrix} \kappa_1(k) & 0 \\ 0 & \lambda_1(k) \end{bmatrix}, \quad q_2(k) = \begin{bmatrix} \kappa_2(k) & 0 \\ 0 & \lambda_2(k) \end{bmatrix}.$$

$q_1(k)$ is confidence matrix of the driver and $q_2(k)$ is the confidence matrix of the active safety system. $\kappa_1(k)$ and $\kappa_2(k)$ are the weight factors, which are related to the driving privilege assignment weight. $\lambda_1(k)$ and $\lambda_2(k)$ are the dynamic factors, which are related to the dynamic characteristics of the driving privilege assignment. Both $Q_1(k)$ and $Q_2(k)$ are semi-positive definite matrixes. And both $R_1$ and $R_2$ are positive definite matrixes. $T_1(k)$ and $T_2(k)$ are the local target trajectories of the 2 players. $T_1(k)$ and $T_2(k)$ need to be updated before each non-cooperative MPC optimization. The update equations are

$$\begin{aligned} T_1(k+1) &= GT_1(k) + Ht_1(k+1) \\ T_2(k+1) &= GT_2(k) + Ht_2(k+1) \end{aligned} \tag{18}$$

With

$$G = \begin{bmatrix} 0 & I_j & 0 & 0 & \cdots & 0 \\ 0 & 0 & I_j & 0 & \cdots & 0 \\ 0 & 0 & 0 & I_j & \cdots & 0 \\ \vdots & \vdots & \vdots & \vdots & \ddots & \vdots \\ 0 & 0 & 0 & 0 & \cdots & I_j \\ 0 & 0 & 0 & 0 & \cdots & 0 \end{bmatrix}, \quad H = \begin{bmatrix} 0 \\ 0 \\ 0 \\ \vdots \\ 0 \\ I_j \end{bmatrix}.$$

Where $I_j$ is $j$-dimensional unit matrix. And $j$ is the number of state variables.

For the lateral system model, $t_1(k)$ and $t_2(k)$ are

$$t_1(k) = \begin{bmatrix} d_{y,D}(k) \\ \psi_D(k) \end{bmatrix}, \quad t_2(k) = \begin{bmatrix} d_{y,A}(k) \\ \psi_A(k) \end{bmatrix}$$

Where $d_{y,D}(k)$ and $\psi_D(k)$ are the desired lateral displacement and the desired yaw angle of the driver. $d_{y,A}(k)$ and $\psi_A(k)$ are the desired lateral displacement and the desired yaw angle of the active safety system.

For the longitudinal system model, $t_1(k)$ and $t_2(k)$ are

$$t_1(k) = \begin{bmatrix} d_{x,D}(k) \\ v_{x,D}(k) \end{bmatrix}, \quad t_2(k) = \begin{bmatrix} d_{x,A}(k) \\ v_{x,A}(k) \end{bmatrix}$$

Where $d_{x,D}(k)$ and $v_{x,D}(k)$ are the desired relative distance and the desired longitudinal velocity of the driver. $d_{x,A}(k)$ and $v_{x,A}(k)$ are the desired relative distance and the desired longitudinal velocity of the active safety system.

4.3 Nash Equilibrium

Two error variables are defined as

$$\begin{aligned} \varepsilon_1(k) &= T_1(k) - \Psi x(k) - \Theta_2 U_2(k) - \Xi D(k) \\ \varepsilon_2(k) &= T_2(k) - \Psi x(k) - \Theta_1 U_1(k) - \Xi D(k) \end{aligned} \tag{19}$$

The cost functions can be transformed according to the error variables

$$V_i(k) = \|\boldsymbol{\Theta}_i U_i(k) - \boldsymbol{\varepsilon}_i(k)\|^2_{Q_i(k)} + \|U_i(k)\|^2_{R_i}, \ i=1,2 \tag{20}$$

The partial derivative of $V_i(k)$ to $U_i(k)$ is

$$\frac{\partial V_i(k)}{\partial U_i(k)} = -2\boldsymbol{\Theta}_i^T Q_i(k)\boldsymbol{\varepsilon}_i(k) + 2[\boldsymbol{\Theta}_i^T Q_i(k)\boldsymbol{\Theta}_i + R_i]U_i(k), \ i=1,2 \tag{21}$$

Since $Q_i(k)$ is semi-definite matrix and $R_i$ is positive definite matrix, the second-order partial derivative is always larger than 0. Therefore, the solution of the first-order partial derivative equal to 0 is the minimum value of the cost function of the $i$-th player, i.e.

$$\begin{aligned} U_i^o(k) &= F_i(k)\boldsymbol{\varepsilon}_i(k) \\ U_i^o(k) &\in \mathbb{U}_i^o, \ i=1,2 \end{aligned} \tag{22}$$

With

$$F_i(k) = [\boldsymbol{\Theta}_i^T Q_i(k)\boldsymbol{\Theta}_i + R_i]^{-1}\boldsymbol{\Theta}_i^T Q_i(k)$$

Where $U_i^o(k)$ represents the optimal control sequence that minimizes the cost function of the $i$-th player. $\mathbb{U}_i^o$ is the set of all optimal control of the $i$-th player.

The non-cooperative MPC is solved by using a iterative method in previous studies[23, 32]. The following theorem shows that the Nash equilibrium solution for the non-cooperative MPC can be achieved by a non-iterative method.

**Theorem 1**: For the system model defined in (11) and the cost function defined in (16), the dynamic game has a unique Nash equilibrium solution if and only if $I-L(k)$ is reversible. And the Nash equilibrium solution of the non-cooperative MPC is

$$\begin{bmatrix} U_1^*(k) \\ U_2^*(k) \end{bmatrix} = K(k)\left\{ \begin{bmatrix} T_1(k) \\ T_2(k) \end{bmatrix} - \begin{bmatrix} \boldsymbol{\Psi} & \boldsymbol{\Xi} \\ \boldsymbol{\Psi} & \boldsymbol{\Xi} \end{bmatrix} \begin{bmatrix} x(k) \\ D(k) \end{bmatrix} \right\}$$

With

$$K(k) = [I - L(k)]^{-1}M(k), \quad M(k) = \begin{bmatrix} F_1(k) & 0 \\ 0 & F_2(k) \end{bmatrix}, \quad L(k) = \begin{bmatrix} 0 & -F_1(k)\boldsymbol{\Theta}_2 \\ -F_2(k)\boldsymbol{\Theta}_1 & 0 \end{bmatrix}.$$

**Proof:**

(1) Existence and uniqueness. In non-cooperative MPC, all the players only know the initial state $x(1)$ in each optimization step. And the remaining states $x(2),\dots,x(N_p)$ are state predictions. The information set of the $i$-th player is $\eta_i(k)=\{x(1)\}$. Hence, the non-cooperative MPC is an open-loop dynamic game. When the system model is linear and the cost function is quadratic, the dynamic game is a linear quadratic game. The existence and uniqueness of the Nash equilibrium solution of the open-loop linear quadratic game can be discriminated by the reversibility of a predefined matrix[33]. Hence, the non-cooperative MPC has a unique Nash equilibrium solution if and only if $P(k)$ is reversible. Where

$$P(k) = \begin{bmatrix} \boldsymbol{\Theta}_1^T Q_1(k)\boldsymbol{\Theta}_1 + R_1 & \boldsymbol{\Theta}_1^T Q_1(k)\boldsymbol{\Theta}_2 \\ \boldsymbol{\Theta}_2^T Q_2(k)\boldsymbol{\Theta}_1 & \boldsymbol{\Theta}_2^T Q_2(k)\boldsymbol{\Theta}_2 + R_2 \end{bmatrix}$$

The $P(k)$ can be transformed into

$$P(k) = \begin{bmatrix} \boldsymbol{\Theta}_1^T Q_1(k)\boldsymbol{\Theta}_1 + R_1 & 0 \\ 0 & \boldsymbol{\Theta}_2^T Q_2(k)\boldsymbol{\Theta}_2 + R_2 \end{bmatrix}[I - L(k)]$$

Since $Q_i(k)$ is a semi-definite matrix and $R_i$ is a positive definite matrix, $P(k)$ is reversible if and only if $I$-$L(k)$ is reversible. Hence, the non-cooperative MPC has a unique Nash equilibrium solution if and only if $I$-$L(k)$ is reversible.

(2) Construction. The optimal response function of the non-cooperative MPC can be represented in a matrix form as

$$\begin{bmatrix} U_1^o(k) \\ U_2^o(k) \end{bmatrix} = M(k)\left\{ \begin{bmatrix} T_1(k) \\ T_2(k) \end{bmatrix} - \begin{bmatrix} \Psi & \Xi \\ \Psi & \Xi \end{bmatrix} \begin{bmatrix} x(k) \\ D(k) \end{bmatrix} \right\} + L(k) \begin{bmatrix} U_1(k) \\ U_2(k) \end{bmatrix} \tag{23}$$

When $I$-$L(k)$ is reversible, the non-cooperative MPC has a unique Nash equilibrium solution. The Nash equilibrium solution is denoted as $(U_1^*(k), U_2^*(k))$. The Nash equilibrium solution should satisfy the optimal response function, i.e.

$$\begin{bmatrix} U_1^*(k) \\ U_2^*(k) \end{bmatrix} = M(k)\left\{ \begin{bmatrix} T_1(k) \\ T_2(k) \end{bmatrix} - \begin{bmatrix} \Psi & \Xi \\ \Psi & \Xi \end{bmatrix} \begin{bmatrix} x(k) \\ D(k) \end{bmatrix} \right\} + L(k) \begin{bmatrix} U_1^*(k) \\ U_2^*(k) \end{bmatrix} \tag{24}$$

If $I$-$L(k)$ is reversible, (24) has a solution. The Nash equilibrium solution of the non-cooperative MPC is

$$\begin{bmatrix} U_1^*(k) \\ U_2^*(k) \end{bmatrix} = [I - L(k)]^{-1} M(k)\left\{ \begin{bmatrix} T_1(k) \\ T_2(k) \end{bmatrix} - \begin{bmatrix} \Psi & \Xi \\ \Psi & \Xi \end{bmatrix} \begin{bmatrix} x(k) \\ D(k) \end{bmatrix} \right\} \tag{25}$$

This completes the proof. □

In MPC, a local optimal solution within the preview horizon is solved at each step. And the preview horizon will recede after each optimization. Only the first control input in each step works. Therefore, the feedback gains of the 2 players in the non-cooperative MPC are

$$\begin{bmatrix} K_1(k) \\ K_2(k) \end{bmatrix} = \begin{bmatrix} I_l & 0 & \cdots & 0 & 0 & 0 & \cdots & 0 \\ 0 & 0 & \cdots & 0 & I_l & 0 & \cdots & 0 \end{bmatrix} K(k) \tag{26}$$

$$\underbrace{\phantom{I_l \;\; 0 \;\; \cdots \;\; 0}}_{N_p} \underbrace{\phantom{0 \;\; 0 \;\; \cdots \;\; 0}}_{N_p}$$

Where $I_l$ is $l$-dimensional unit matrix. And $l$ is the number of control variables.

The control inputs of the 2 players (driver and active safety system) in the non-cooperative MPC can be expressed as

$$\begin{aligned} u_D^*(k) &= K_1(k)\left\{ \begin{bmatrix} T_1(k) \\ T_2(k) \end{bmatrix} - \begin{bmatrix} \Psi & \Xi \\ \Psi & \Xi \end{bmatrix} \begin{bmatrix} x(k) \\ D(k) \end{bmatrix} \right\} \\ u_A^*(k) &= K_2(k)\left\{ \begin{bmatrix} T_1(k) \\ T_2(k) \end{bmatrix} - \begin{bmatrix} \Psi & \Xi \\ \Psi & \Xi \end{bmatrix} \begin{bmatrix} x(k) \\ D(k) \end{bmatrix} \right\} \end{aligned} \tag{27}$$

The Nash equilibrium inputs of the driver and the active safety system in the driving privilege gradual handover strategy is obtained by using (27). Although the process of solving the optimal solution by using (22) is convenient and clear, the calculation result of (22) may be numerical instability. Therefore, the method introduced in [34] is applied to avoid the numerical instability of $F_i(k)$, i.e.

$$\begin{aligned} F_1(k) &= \begin{bmatrix} S_{Q_1}(k)\Theta_1 \\ S_{R_1} \end{bmatrix}^+ \begin{bmatrix} S_{Q_1}(k) \\ 0 \end{bmatrix} \\ F_2(k) &= \begin{bmatrix} S_{Q_2}(k)\Theta_2 \\ S_{R_2} \end{bmatrix}^+ \begin{bmatrix} S_{Q_2}(k) \\ 0 \end{bmatrix} \\ &\begin{cases} S_{Q_1}(k)^T S_{Q_1}(k) = Q_1(k) \\ S_{R_1}{}^T S_{R_1} = R_1 \end{cases} \\ &\begin{cases} S_{Q_2}(k)^T S_{Q_2}(k) = Q_2(k) \\ S_{R_2}{}^T S_{R_2} = R_2 \end{cases} \end{aligned} \tag{28}$$

Where $A^+$ is the generalized inverse matrix of $A$.

## 5. Simulation Verification

5.1 Parameter Specification

The vehicle parameters are shown in Table 4. The preview horizon is chosen as $N_p$=10, and the control horizon is chosen as $N_u$=10. The time step is $T$=0.01s. Similar to optimal control, only the relative values of $q_1(k)$, $q_2(k)$ and $r_1$, $r_2$ have influence on the control result in non-cooperative MPC. Therefore, $r_1$=1 and $r_2$=1 are set in the simulation.

Table 4 Vehicle parameters

| Symbol | Unit | Value |
| --- | --- | --- |
| $a$ | m | 1.0 |
| $b$ | m | 1.5 |
| $m$ | kg | 1270 |
| $I_z$ | kg·m² | 1443.1 |
| $C_f$ | kN/rad | 30 |
| $C_r$ | kN/rad | 30 |

5.2 Lane-change Scenario

Firstly, the gradual takeover strategy is verified by using the lane-change scenario. The driver of the host vehicle desires a left lane-change, but the driver does notice a target vehicle in the rear of the target line is approaching. The active safety system considers that it is not suitable for lane-change after risk assessment, and the active safety system expects lane keeping. There is a conflict between the intention of the driver and active safety system. The gradual takeover strategy carries out a driving privilege handover at the appropriate time based on the result of risk assessment.

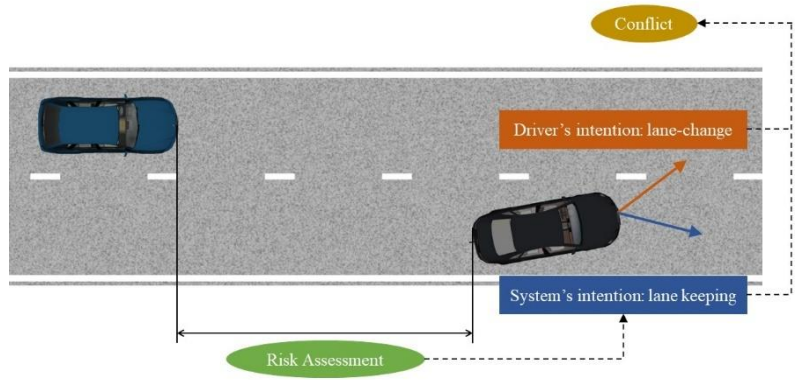

Fig 17. The conflict in lane-change scenario

The driver desire to change the lane at 2s. The length of the lane-change trajectory is 80m and the width is 3.5m. The lane-change trajectory uses a fifth-order polynomial trajectory[35]. The velocity of the host vehicle is 20m/s. In lane-change case 1, the target vehicle is 30m behind, and the velocity of the target vehicle is 23m/s. In lane-change case 2, the target vehicle is 40m behind, and the velocity of the target vehicle is 23m/s. $\lambda_1(k)$ and $\lambda_2(k)$ are set to be constant in these 2 cases, i.e. $\lambda_1(k)$=2, $\lambda_2(k)$=2. In lane-change case 1, the risk level change from RL0 to RL2 at 3.2s. $\kappa_1(k)$=0.1 and $\kappa_2(k)$=0 at the beginning. At 3.2s, $\kappa_1(k)$ linearly decreases to 0 and $\kappa_2(k)$ linearly increases to 0.1 within 1s. In lane-change case 2, the risk level change from RL0 to RL2 at 4.9s. $\kappa_1(k)$=0.1 and $\kappa_2(k)$=0 at the beginning. At 4.9s, $\kappa_1(k)$ linearly decreases to 0 and $\kappa_2(k)$ linearly increases to 0.1 within 1s.

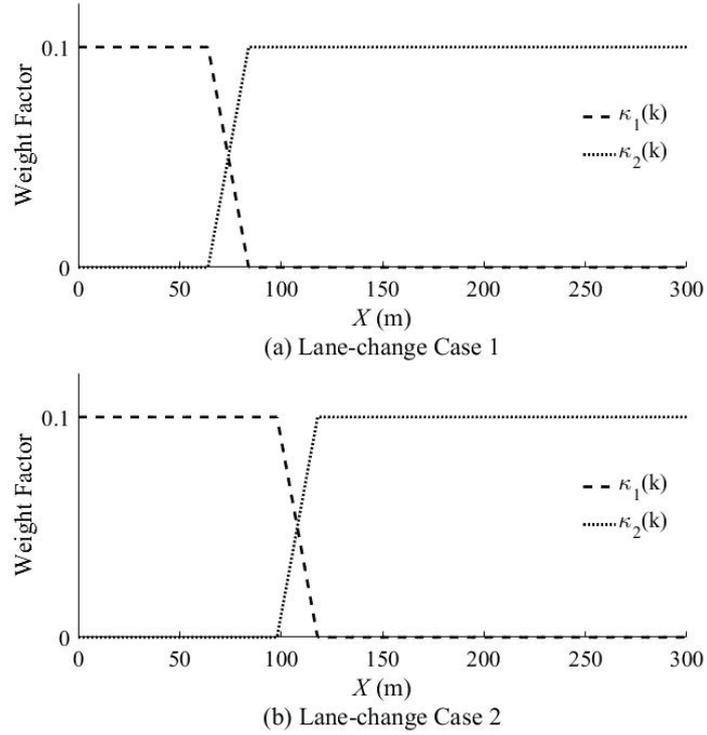

Fig 18. The weight factor in the lane-change scenario

The $\kappa_1(k)$ and $\kappa_2(k)$ in these 2 cases are shown in Fig 18. The simulation results are shown in Fig 19. When the driving privilege is transformed during the lane-change, the vehicle can smoothly return to the initial lane. The driving privilege gradual handover strategy can realize the gradual transition of the driving privilege between the driver and active safety system when the risk level is raised, so that the vehicle returns to the initial lane. The gradual takeover strategy can assist the driver to avoid the danger and can ensure that the intervention of the active safety system is not too abrupt. A stable trajectory can be planned in these 2 cases. The front wheel angle is kept in a small range, which is very beneficial for maintaining vehicle stability.

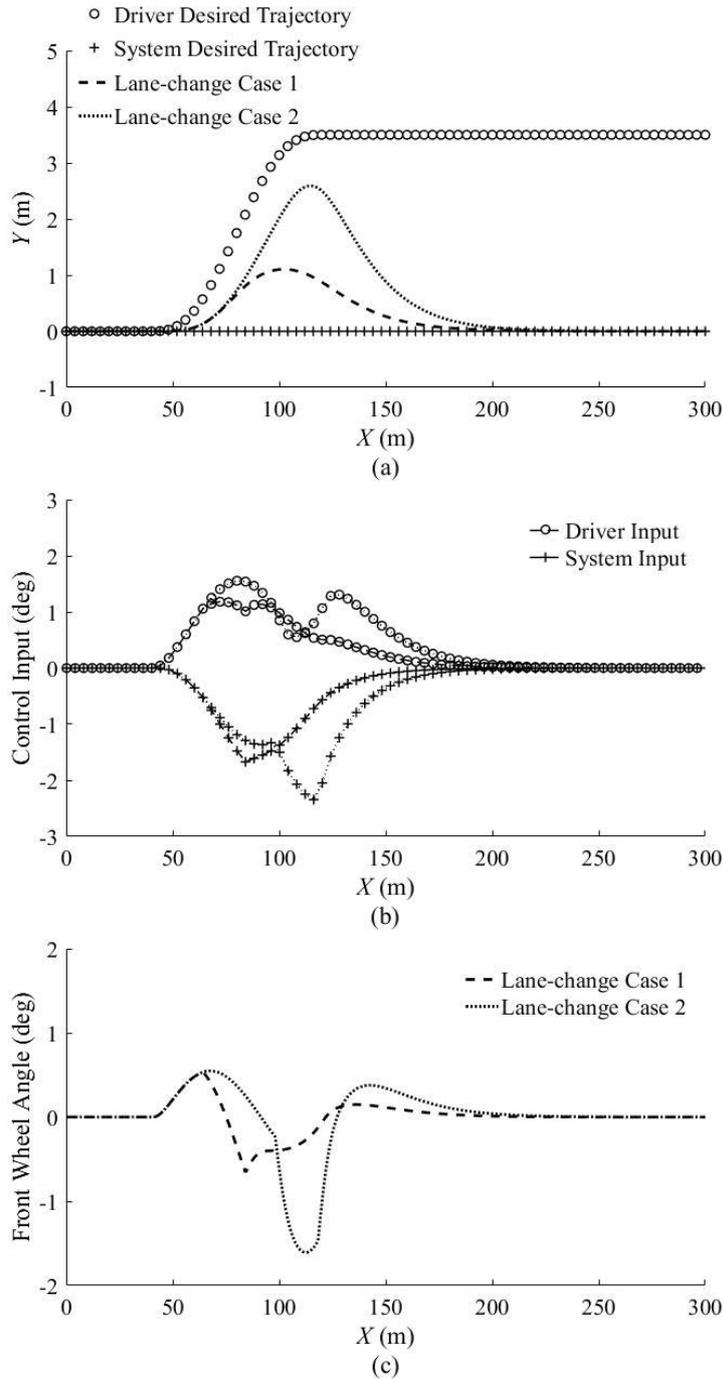

Fig 19. The simulation results of the lane-change scenario

### 5.3 Cut-in Scenario

Secondly, the gradual takeover strategy is verified by using the cut-in scenario. When the host vehicle goes straight, the target vehicle in the lane beside cuts in. The driver of the host vehicle did not notice the cut-in of the target vehicle, and no action was taken. Hence, the driver's intention is velocity keeping. The active safety system starts the risk assessment after the target vehicle crosses the lane line. And the active safety system decides to decelerate to the same velocity of the target vehicle based on the result of risk assessment. The conflict arises at this time. The gradual takeover strategy transfer the driving privilege from the driving to the active safety system based on the result of risk assessment.

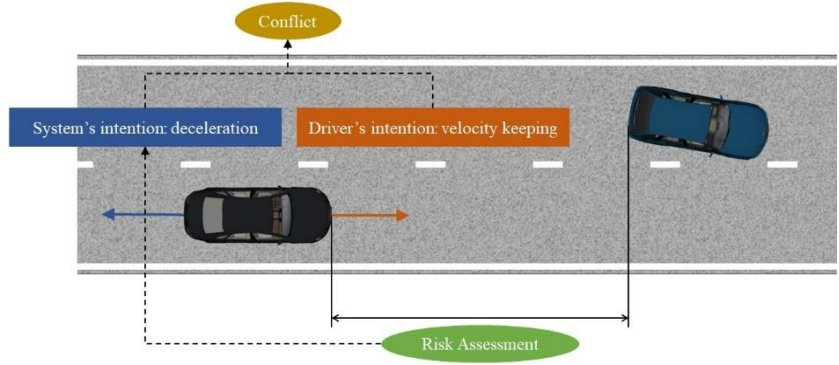

Fig 20. The conflict in cut-in scenario

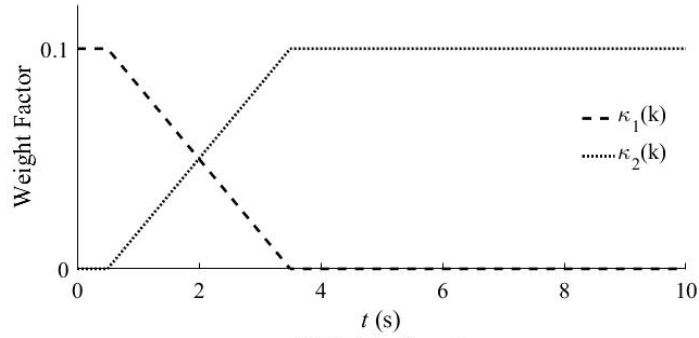

(a) Cut-in Case 1

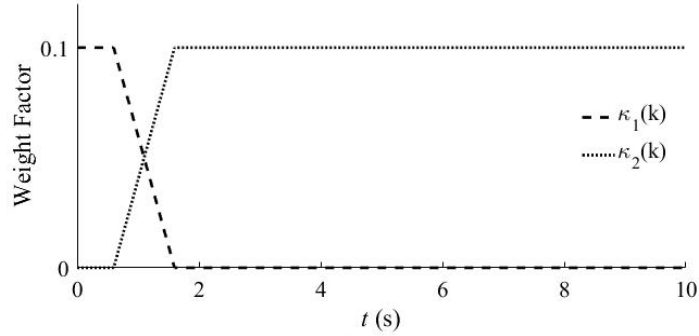

(b) Cut-in Case 2

Fig 21. The weight factor in the cut-in scenario

In cut-in case 1, the initial velocity of the host vehicle is 8m/s. The target vehicle starts to cut in 10m before the host vehicle, and the velocity of the target is 5m/s. In cut-in case 2, the initial velocity of the host vehicle is 12m/s. the target vehicle starts to cut in 10m before the host vehicle, and the velocity of the target is 10m/s. $\lambda_1(k)$ and $\lambda_2(k)$ are set to be constant in these 2 cases, i.e. $\lambda_1(k)=100$, $\lambda_2(k)=100$. In cut-in case 1, the risk level change from RL0 to RL1 at 0.5s. $\kappa_1(k)=0.1$ and $\kappa_2(k)=0$ at the beginning. At 0.5s, $\kappa_1(k)$ linearly decreases to 0 and $\kappa_2(k)$ linearly increases to 0.1 within 3s. In cut-in case 2, the risk level change from RL0 to RL2 at 0.6s. $\kappa_1(k)=0.1$ and $\kappa_2(k)=0$ at the beginning. At 0.6s, $\kappa_1(k)$ linearly decreases to 0 and $\kappa_2(k)$ linearly increases to 0.1 within 1s.

The $\kappa_1(k)$ and $\kappa_2(k)$ in these 2 cases are shown in Fig 21. The simulation results are shown in Fig 22. When the target vehicle cuts in, the risk level increases as the relative distance decreases. The gradual takeover strategy begins to transfer the driving privilege from the driver to the active safety system when the risk level reaches the corresponding threshold. The brake deceleration is small when the risk level is low. The comfort is satisfied while the safety is ensured. When the risk level is high, the brake deceleration is increased to ensure safety. The active safety system in the gradual takeover strategy can gradually intervene, and the intervention strategy can be adjusted according to the risk level. Therefore, the gradual takeover strategy can better balance comfort and safety.

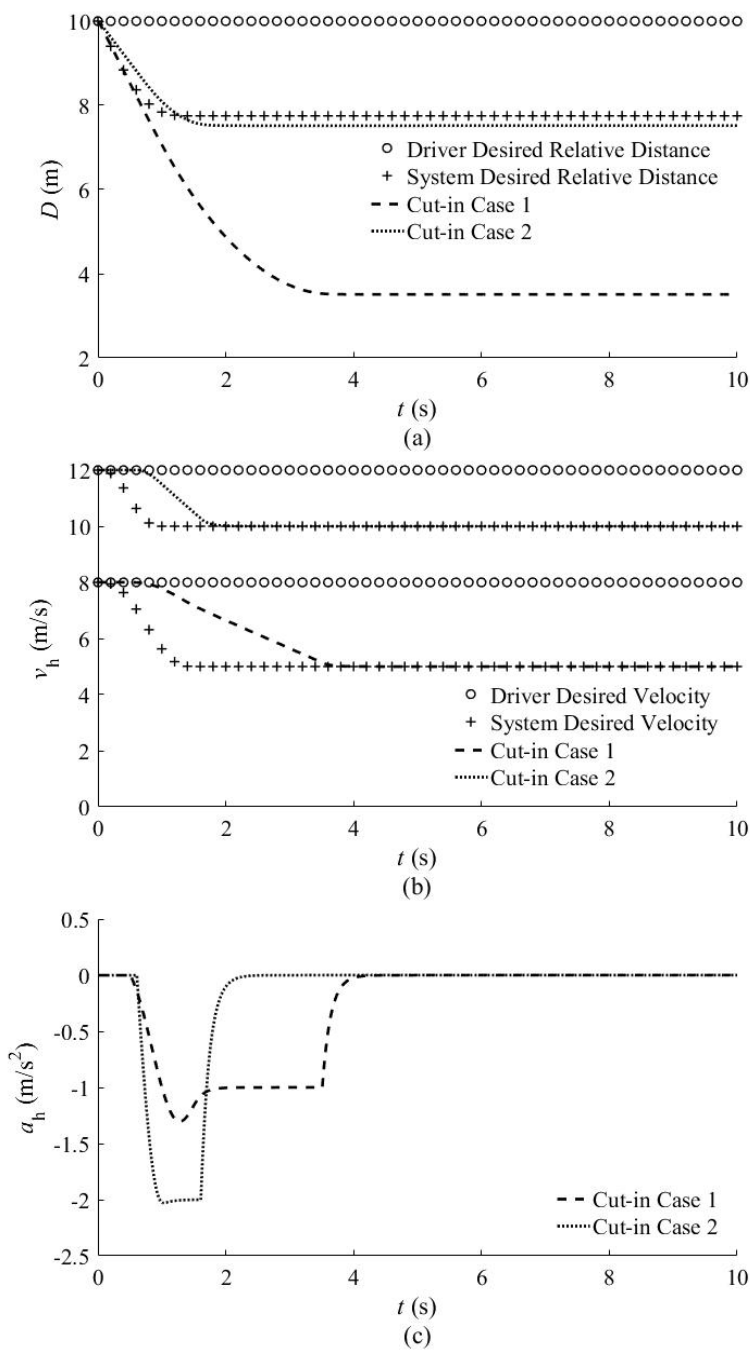

Fig 22. The simulation results of the cut-in scenario

## 6. Conclusions

A gradual takeover strategy of the active safety system is proposed. The driving privilege assignment in real-time is achieved based on the risk level, whereas the driving privilege gradual handover is realized by using the dynamic game. Since the risk level is related to the subjective perception of the driver, the naturalistic driving data is used to study the behavior characteristics of the driver in typical dangerous scenarios. The dangerous and normal car-following cases and cut-in cases are extracted by using the naturalistic driving data. TTC is defined as the obvious risk measure, whereas the reaction time left to the driver if the target vehicle starts to brake is defined as the potential risk measure, i.e. time margin (TM). A risk assessment algorithm is proposed based on the obvious risk and potential risk. The dangerous and normal car-following cases are applied to verify the effectiveness of the risk assessment algorithm. It is identified that the risk assessment algorithm performs better than TTC in ROC. The braking moment of the driver in the cut-in scenario is studied by using the dangerous and normal cut-in cases. The results show that most drivers start braking when the target vehicle reaches the lane line. Therefore, the moment when the target vehicle reaches the lane line is taken as the time that the risk estimation starts in the cut-in scenario. To avoid disturbing the operation of the driver and the difficulties in takeover, the driving privilege gradual handover is realized. The vehicle is jointly controlled by the driver and active safety system during the driving privilege gradual handover. The non-cooperative MPC are employed to deal with conflicts between the driver and active safety system. The system model and cost function are constructed, and the Nash equilibrium solution of non-cooperative MPC is obtained. The driving privilege gradual handover is realized through the update of the confidence matrix. The simulation verification shows that the gradual takeover strategy can realize the driving privilege gradual handover in the dangerous process. The safety can be ensured while the comfortable is maintained.

**Acknowledge**

This work is supported by the national key research and development plan of China (2016YFB0100904-2, 2018YFB1600701), Fok yingdong young teachers fund project (171103), Key Research and Development Program of Shaanxi (2018ZDCXL-GY-05-03-01), and Youth Program of National Natural Science Foundation of China (52002034).

The authors would like to give the sincere gratitude to Zhiwei Feng, Jingwei He and Lan Xia. Zhiwei Feng help to screen the cut-in cases. Jingwei He help to check the dangerous case and select the dangerous case example. Lan Xia help to screen the normal car-following cases.

**Biographies:**

Rui Liu received the B.S. and the M.S. degree in vehicle engineering from Chang'an University, Xi'an, China, in 2012 and 2015, respectively. And he received the Ph.D. degree in vehicle engineering from Tongji University, Shanghai, China, in 2020. He is currently a lecturer with the School of Automobile, Chang'an University, Xi'an, China. His research interest includes vehicle active safety, intelligent vehicle control, naturalistic driving studies, evaluation of intelligent vehicle.

Xichan Zhu received the Ph.D. degree in vehicle engineering from Tsinghua University, Beijing, China, in 1995. From 1996 to 2005, he was worked with China Automotive Technology & Research Center, Tianjin, China. He is currently the professor of School of Automotive Studies, Tongji University, Shanghai, China. His research interest includes vehicle active safety and vehicle passive safety.

Xuan Zhao received the B.S., M.S., and Ph.D. degrees in vehicle engineering from Chang'an University, China, in 2007, 2009, and 2012, respectively, where he is currently a professor with the School of Automobile. He has undertaken over 9 government sponsored works, including the National Key Research and Development Program of China and the China Postdoctoral Science Foundation. His main research interests include intelligent vehicle perception and control, intelligent vehicle evaluation, and electric vehicle control strategy.

Jian Ma received the Ph.D. degree in transportation engineering from Chang'an University, China, in 2001. He is currently a professor with the School of Automobile, Chang'an University. He has undertaken more than 30 government-sponsored research projects, such as 863 projects and key transportation projects in China. He has published more than 90 academic articles, and he has authored 4 books. His main research interests include vehicle dynamics, electric vehicle and clean energy vehicle technology, and automobile detection technology and theory.